\documentclass{emulateapj}
\usepackage[dvips]{color}
\usepackage{amssymb, amsmath, amsbsy, epsfig, epsf, slashbox}

\newcommand{\ledd}{\ensuremath{L\mathrm{_{Edd}}}}
\newcommand{\lratio}{\ensuremath{L/\ledd}}

\newcommand{\lfive}{\ensuremath{\lambda L_{\lambda}(5100)}}

\newcommand{\mbh}{\ensuremath{M_\mathrm{BH}}}

\newcommand{\rs}{\ensuremath{r_{\rm \scriptscriptstyle S}}}

\newcommand{\ha}{{\rm H\ensuremath{\alpha}}}
\newcommand{\hb}{{\rm H\ensuremath{\beta}}}

\newcommand{\nha}{H\ensuremath{\alpha ^{n}}}
\newcommand{\nhal}{H\ensuremath{\alpha ^{n}}}
\newcommand{\bha}{H\ensuremath{\alpha ^{b}}}
\newcommand{\bhb}{H\ensuremath{\beta ^{b}}}

\newcommand{\nhbl}{H\ensuremath{\beta ^{n}}}

\newcommand{\civl}{C\,{\footnotesize IV} $\lambda$1549}
\newcommand{\nvl}{N\,{\footnotesize V} $\lambda$1240}
\newcommand{\nevl}{[Ne\,{\footnotesize V}] $\lambda$3425 }
\newcommand{\oiil}{[O\,{\footnotesize II}] $\lambda$3727}
\newcommand{\neiiil}{[Ne\,{\footnotesize III}] $\lambda$3870 }
\newcommand{\oiiil}{[O\,{\footnotesize III}] $\lambda$5007}
\newcommand{\oil}{{\rm [O\,{\footnotesize I}]}$\lambda$6300}
\newcommand{\niil}{[N\,{\footnotesize II}] $\lambda$6583}
\newcommand{\niir}{[N\,{\footnotesize II}] $\lambda$6548}
\newcommand{\siil}{[S\,{\footnotesize II}] $\lambda \lambda$6717, 6731}

\newcommand{\nevt}{[Ne\,{\tiny V}] $\lambda$3425 }
\newcommand{\oiit}{[O\,{\tiny II}] $\lambda$3727}
\newcommand{\neiiit}{[Ne\,{\tiny III}] $\lambda$3870 }
\newcommand{\oiiit}{[O\,{\tiny III}] $\lambda$5007}
\newcommand{\oit}{{\rm [O\,{\tiny I}]}$\lambda$6300}
\newcommand{\niit}{[N\,{\tiny II}] $\lambda$6583}
\newcommand{\siit}{[S\,{\tiny II}] $\lambda \lambda$6717, 6731}

\newcommand{\oiii}{[O\,{\footnotesize III}]}

\newcommand{\civ}{C\,{\footnotesize IV}}
\newcommand{\niii}{N\,{\footnotesize III}}
\newcommand{\niv}{N\,{\footnotesize IV}}

\newcommand{\nev}{Ne\,{\footnotesize V}}

\newcommand{\neii}{Ne\,{\footnotesize II}}
\newcommand{\neiii}{Ne\,{\footnotesize III}}

\newcommand{\nii}{[N\,{\footnotesize II}]}
\newcommand{\sii}{[S\,{\footnotesize II}]}

\newcommand{\feii}{Fe\,{\footnotesize II}}

\newcommand{\mgii}{Mg\,{\footnotesize II}}

\def\lax{{$\mathrel{\hbox{\rlap{\hbox{\lower4pt\hbox{$\sim$}}}\hbox{$<$}}}$}}
\def\gax{{$\mathrel{\hbox{\rlap{\hbox{\lower4pt\hbox{$\sim$}}}\hbox{$>$}}}$}}

\begin{document}


\title{ THE BALDWIN EFFECT IN THE NARROW EMISSION LINES OF AGNS}

\author{
Kai~Zhang\altaffilmark{1,2}, Ting-Gui Wang\altaffilmark{1}, C.~Martin~Gaskell\altaffilmark{3}, and Xiao-Bo~Dong\altaffilmark{1}}

\altaffiltext{1}{Key Laboratory for Research in Galaxies and
Cosmology, The University of Sciences and Technology of China,
Chinese Academy of Sciences, Hefei, Anhui 230026, China; \\
zkdtc@mail.ustc.edu.cn; twang@ustc.edu.cn; xbdong@ustc.edu.cn  }

\altaffiltext{2}{Key Laboratory for Research in Galaxies and Cosmology, Shanghai
Astronomical Observatory, Chinese Academy of Sciences, 80 Nandan
Road, Shanghai 200030, China}

\altaffiltext{3}{Centro de Astrof\'isica de Valpara\'iso y Departamento de F\'isica y Astronom\'ia, Facultad de Ciencias,
Universidad de Valpara\'iso, Av. Gran Breta\~na 1111, Valpara\'iso, Chile.
martin.gaskell@uv.cl}

\email{zkdtc@mail.ustc.edu.cn}
\shorttitle{Narrow Line BE}
\shortauthors{Zhang et al.}

\begin{abstract}

The anti-correlations between the equivalent widths of emission lines and the
continuum luminosity in AGNs, known as the Baldwin effect
are well established for broad lines, but are less well studied for
narrow lines. In this paper we explore the Baldwin effect of narrow emission lines over a wide
range of ionization levels and critical densities using a large sample
of broad-line, radio-quiet AGNs taken from Sloan Digital Sky Survey (SDSS)
Data Release 4. These type1 AGNs span three orders of magnitude in continuum luminosity.
We show that most narrow lines show a similar Baldwin effect slope of about -0.2
while the significant deviations of the slopes for \niil, \oiil, \nevl, and the
narrow component of \ha\ can be
explained by the influence of metallicity, star-formation contamination and possibly by difference in the
shape of the UV-optical continuum. The slopes do not show any correlation with
either the ionization potential or the critical density. We
show that a combination of 50\% variations in continuum near 5100\AA\ and a log-normal
distribution of observed luminosity can naturally reproduce a constant
Baldwin effect slope of -0.2 for all narrow lines. The variations of the continuum
could be due to variability, intrinsic anisotropic emission, or an inclination
effect.

\end{abstract}

\keywords{galaxies: active--galaxies:Seyfert--(galaxies:) quasars:
emission lines }

\section{Introduction}
The anti-correlation between the equivalent widths of broad emission lines and
AGN luminosity (the ``Baldwin effect''; hereinafter BE) was first discovered by Baldwin (1977) for the \civl\ broad
emission line in high redshift AGNs. It was initially hoped to be able to use the
effect to calibrate the AGN luminosity to be able to use AGNs as cosmological
standard candles (Baldwin et al. 1978), but the large dispersion of this relationship
rendered this impossible (Baldwin et al. 1989; Zamorani et al. 1992). The BE is now
well established for nearly all broad emission lines and the slope of the BE
steepens with increasing ionization potential (Zheng \& Malkan 1993;
Dietrich et al. 2002). Several mechanisms have been proposed to explain this
effect (See Shields 2007 for a review). Among these perhaps the most widely accepted one
is that the ionizing continuum softens with increasing luminosity so
there are relatively fewer ionizing photons for broad emission line
formation in high-luminosity AGNs.
This model can reproduce the ionization
energy--BE slope relationship fairly well (Korista et al. 1998) and the
assumption has observational support (Binette et al.
1989; Zheng \& Malkan 1993; Wang \& Lu 1998; Korista et al. 1998).
Some theoretical models (Netzer 1985, 1987; Netzer, Laor, \& Gondhalekar 1992; Wandel 1999a,b)
could produce a softer ionizing spectrum in high-luminosity sources,
but the standard thin disc model (Shakura \& Sunyaev 1973)
they adopt suffers from many problems (Antonucci 2002; Gaskell \& Klimek 2003;
Gaskell 2008; Lawrence 2012; Antonucci 2012).
In this sense, the explanation of broad line BE is still elusive and
controversial. An outlier of the ionization energy--BE
slope relationship is \nvl, which has an ionization energy of
97.7eV but shows no BE at all (Dietrich et al. 2002). It has been proposed that this can be explained
by a dependence of metallicity on AGN luminosity, which in turn is a
combination of Eddington ratio (\lratio) and black hole mass (Korista et
al. 1998; Hamann \& Ferland 1993, 1999; Dietrich et al. 1999;
Dietrich \& Wilhelm-Erkens 2000). 
In local galaxies, gas metallicity correlates well with the mass of
galaxies (Tremonti et al. 2004), and more massive galaxies have larger
black hole masses (\mbh) according to the \mbh-$M_*$ relationship.
For Type1 AGN population, the brighter AGNs would have higher \mbh\
on average (Kollmeier et al. 2006; Steinhardt \& Elvis 2010a,b; Lusso et al. 2012)
and thus higher metallicities.
Recent studies of the BE for \civ, \mgii\ and \feii, however,
challenge this picture by showing that
the BE might instead be driven by the EW's correlation
with Eddington ratio (Baskin \& Laor 2004; Bachev et al.
2004; Warner, Hamann \& Dietrich 2004; Zhou et al. 2006; Dong et al. 2009a,b)
or with \mbh (Netzer, Laor \& Gondhalekar 1992; Wandel,
Peterson \& Malkan 1999; Shields 2007; Kovacevic et al. 2011).

The BE of narrow emission lines is much less
well studied, and some results are still controversial. Steiner (1981) discovered a strong BE
for \oiiil\ (especially in AGNS with strong optical \feii) and
Wills et al. (1993) found the narrow lines in high-luminosity AGNs are very weak.
Meanwhile, Croom et al. (2002) found a significant BE for \nevl\ and \oiil\ but obtained a null
result for \neiiil\ and \oiiil\ using the 2dF sample. Subsequent work shows,
however, that \oiiil\ does show a BE (Dietrich et al. 2002, Netzer et al.
2004). H\"onig et al. (2008) and Keremedjiev et al. (2009), used Spitzer
data to find that the BEs of different mid-IR narrow emission lines have
a nearly constant slope. Several mechanisms have been proposed to explain the NLR BE. NLRs follow a
size--luminosity relationship (Schmitt et al. 2003b; Bennert et al.
2002; Greene et al. 2011). The size of the NLR may grow beyond
the size of the host galaxy in high-luminosity AGNs,  thus
turning from ionization-bounded to matter-bounded, and so producing
the BE (Croom et al. 2002). As the EW is proportional to the
covering factor (CF) and it is found that the CF contributes
to much of the variance of the EW (Baskin \& Laor 2005), a luminosity-
dependent CF is also a possible cause of the BE (Shields et al. 1995; Stern \& Laor 2012b;
Stern \& Laor 2012c).
It has recently been proposed that the EW of narrow emission lines is dependent on
the inclination to the accretion disc (Risaliti et al. 2011), at
least in the highest EW sources. In principle, this is a potential
cause of the BE too.

To make progress in understanding the NLR BE, we need to measure narrow
emission lines of a wide range of ionization potentials and critical
densities, and determine the nature of their BEs more accurately. In this paper, we use a
well-defined sample drawn from SDSS DR4 and employ the technique of
composite spectra to investigate the BE for prominent narrow
emission lines in the optical band and to study the origin
of the NLR BE. We use a cosmology with
$H_{\rm 0}$ = 70 km\,s$^{-1}$\,Mpc$^{-1}$, $\Omega_{\rm m}$ = 0.3, and
$\Omega_{\rm \Lambda}$ = 0.7 throughout this paper.

\section{Sample and Measurements}

\subsection{Sample}
We need accurate emission line and continuum measurements
to ensure reliable determinations of equivalent widths and continuum
luminosities.
From the spectral data set of the Sloan Digital Sky
Survey Fourth Data Release (Adelman-McCarthy et al. 2006),
we have selected 4178 Seyfert 1 galaxies and quasars (i.e., type~1 AGNs)
as described in Dong et al. (2011).
We apply a redshift cutoff of 0.8 so that the redshift of
the spectrum can be accurately determined using \oiiil.
To ensure high-quality spectra we
require a median signal-to-noise ratio (S/N) of $\geq 10$ per pixel in the
optical. To minimize the host-galaxy contamination (see the
Appendix of Dong et al. 2011), we restrict the weak stellar absorption features,
such that the rest-frame EWs of Ca\,K
(3934 \AA), Ca\,H + H$\epsilon$ (3970 \AA), and H$\delta$ (4102 \AA) absorption
features are undetected at $< 2\,\sigma$ significance. This criterion ensures
that the host contamination is less than 10\% around 4200\AA (Dong et al. 2011).
One may note that Dong et al. (2011) do not consider very
young stellar population (i.e., emission line galaxies). But previous
analysis have shown that in massive galaxies, the optical continuum is not
dominated by very young stellar population although the UV continuum may be
(Schawinski et al. 2007).
After removing duplications and sources with too
many bad pixels in the \hb\ + \oiii\ region, we obtain 4178 type~1 AGNs.
Including radio-loud AGNs may influence the measurement of EW of emission thus
produce false effect for two reasons: firstly, the jet may interact with the ISM to enhance
the narrow line emission (Labiano et al. 2007) and secondly, it might also
enhance the continuum through beaming if the jet points close to our line of sight.
By matching with the FIRST catalog (Becker et al. 1995) using the method of
Lu et al. (2010), we reject 499 radio-loud AGNs so that our final sample consists of
3677 sources.

\subsection{Spectral fitting and measurements}
We give a brief description of our spectrum fitting process here; the
details can be found in Dong et al. (2011).
To model the spectrum, we used a code based on the MPFIT package (Markwardt 2009)
and fit the AGN featureless continuum, the \feii\ multiplets,
and other emission lines simultaneously. The
AGN continuum is represented locally by a power-law, for the region of 4200--5600
\AA\ and for the \ha\ region (if present).
The \feii\ template by V\'eron-Cetty et al. (2004) that we use, is
constructed using the identification and measurement of \feii\ lines in I~Zw~1.
It has two separate sets of templates in \emph{analytical} forms, one for
the broad-line system and the other for the narrow-line system.
Within each system, the relative velocity shifts and relative strength
are assumed to be the same as those in I~Zw~1. Broad \feii\ lines share
the same profile as broad \hb, while each narrow \feii\ lines
is modeled with a Gaussian. During the fitting, the normalization and redshift
of each system are taken as free parameters.
The broad Balmer lines are
fitted with as many Gaussians as is statistically justified. All narrow emission
lines, except for the \oiii\,$\lambda
\lambda$4959, 5007 doublet lines,  are fitted with a single Gaussian.
Each line of the \oiii\ doublet is modeled with two
Gaussians, one accounting for the line core and the other for a possible blue
wing, as seen in many objects.
Since the sources in our sample do not suffer from significant host galaxy contamination, we do not
apply starlight corrections to individual spectra.
For each source, we use \lfive\ and FWHM of broad \hb\ line to obtain the \mbh\ using
the virial mass estimates by the Dibai method (Dibai 1977) using the formalism of Wang et al. (2009).
The typical statistical scatter about the \mbh\ obtained by reverberation-mapping
is about 0.4~dex, and it may also be subject to more systematic errors. (Krolik 2001; Collin et al. 2006;
Shen et al. 2008; Fine et al. 2008; Marconi et al. 2008; Denney et al. 2009; Rafiee \& Hall 2011a;
Steinhardt 2011). The error of \lratio\ is of similar magnitude as that of \mbh.

\subsection{Composite Spectra Generating and Fitting}
A convenient way to explore the correlations between EWs and
other parameters like luminosity is to
make composite spectra for different parameter bins.
We normalize individual spectrum to the mean flux around 4200\AA\
and then construct the geometric composite as Vanden Berk et al. (2001).
To get an accurate continuum measurement, we need
to subtract the broad emission lines, especially, \ha\ and \hb\ and \feii\
emission from the spectrum. For the broad component of \hb\ and \feii\, we subtract them
from the original spectrum; but for
the broad component of \ha, we leave it un-subtracted when making composite spectra.
This is because the broad \ha\ is highly blended with
\niil\ and \niir, so deblending in individual spectrum is not reliable while
deconvolving the blends in the high SN composite spectrum is easier.
The fitting algorithm used to model the composite spectrum
is the same as described in Section 2.2. More specifically,
we fit the blends using one gaussian for the narrow component of \ha, \niil\ and
\niir, two gaussians for \siil\ and three for \bha.
The broad-line-subtracted spectra are shown in Fig.~1. The embedded panels show the fitting
result for the \bha + \nii\ blends.

\section{Results}

\subsection{The Baldwin effect for different lines}
First, we want to explore if the BE exists in the prominent narrow lines
 \nevl, \oiil, \neiiil, \nhbl, \oiiil, \oil, \nhal, \niil\
and \siil. Measuring narrow lines in individual spectrum are subject to the
S$/$N limit and the EW-$L_{5100}$ relation of all our lines shows a large dispersion
(typical 0.2~dex, this can be seen clearly in the middle panel of Fig.~4),
so we turn to composite spectra for reliable measurements.
We divide our sample into intervals of 0.3~dex in
$5100\AA$ luminosity starting at log $L_{5100}$ = 43.2 to 45.9 $erg s^{-1}$.
and make a composite spectrum in each luminosity bin
as described in Section 2.3.
From Fig.1 we can see two clear and remarkable results:
\begin{itemize}
    \item
    With increasing luminosity, the narrow lines vanish.
    \item
    The slope of the observed continuum become bluer with increasing luminosity.
\end{itemize}
To see the dependencies of the EWs on the luminosity more clearly, in Fig.~2 we plot the log EW that derived from
composite spectra for each luminosity bin for each of the lines
listed above against log $L_{5100}$  in order to get more qualitative results.
For each line, we show a weighted linear regression of the logarithm of the EW on the
logarithm of the luminosity and we list all the
BE slopes of the narrow emission lines together with their ionization
energies and critical densities in Table.1. When fitting the relationship, we add an
error to each data point that would help to reduce the normalized $\chi^2$ to
$\sim$1. These added errors are the same for individual narrow line, and they
account for the potential other systematic errors.
The BE slopes of the high-ionization lines \neiiil\ and \nevl\ are $-0.26\pm0.02$, $-0.31\pm0.015$ respectively which are consistent
with the results of Keremedjiev et al. (2009) who obtained $-0.22\pm0.06$, and $-0.19\pm0.06$ respectively
for the \neiii\ $\lambda$ 15.56$\mu$m, and \nev\ $\lambda$ 14.32$\mu$m lines.
Our \oiiil\ BE slope of $-0.21\pm0.016$ is similar to the values of $\sim -0.2$ found by Kovacevic et al. (2011)
and steeper than the $-0.1\pm0.02$ found by Dietrich et al. (2002). The difference might be due to
different sample we employ. The error of slopes we give here include
only the statistic error of the fitting, but do not include the measuring error and
intrinsic dispersion of the EW, which may reach 0.2~dex typically.

In addition to finding the BE in high-ionization lines, we also find the BE in the
low-ionization lines of \oil\, \siil\ and \oiil.  The narrow recombination line:
\nhbl\ also shows a BE with a similar slope as for the forbidden lines.
For comparison, the {\em broad} \hb\ line shows an inverse BE with a slope
of 0.16 for our sample. This is consistent with Greene \& Ho (2005) and
Croom et al. (2002) who reported $EW(\bhb)\propto L_{5100}^{0.13}$ and
$EW(\bhb)\propto L_{5100}^{0.19}$ respectively. We can see in Fig.~2
that the EW of the narrow component of \ha\ increases with luminosity too. These indicate that the behaviors
of broad lines and narrow lines are different.

The correlation coefficients
between EW(\oiii) and \lfive, z, \mbh, \lratio\ are -0.12, -0.05, -0.01, -0.21, respectively.  These are similar
to the results in Zhang et al. (2011).
For our flux-limited sample, the redshift, luminosity, black hole mass, and
Eddington ratio are correlated with each other. The correlation coefficients between
\lfive\ and $z$, \mbh, \lratio\ are 0.81, 0.67, 0.15 respectively. After controlling for $z$ or \lratio\ in a
partial correlation analysis, EW(\oiii) correlates with \lfive\ with \rs=-0.14 and -0.16.
This indicates that the BE of narrow
lines is not a secondary effect of correlations between the EW and redshift or \lratio\
but an independent phenomenon. We note that as the correlation strength is not strong, one
possibility is that there are other factors that regulate the EW as discussed in Zhang et al. (2011).
Also, measurement error in EW(\oiii) may act to smear the correlation. The measurement error
could be introduced by limited S$/$N and fitting process. In spite
of these uncertainties, the BE of narrow lines does exist and may shed light on
physical process in AGNs as explored in detail below.

\subsection{The ionization energy -- BE slope relationship: no correlation}
A dependence of the BE slope on ionization energy is found for
broad emission lines in AGNs (see introduction) and this is the most compelling evidence
for a luminosity-dependent ionizing spectrum.
The NLR lies far from the nucleus, has a complex geometry and may contain dust (Netzer
\& Laor 1993; Tomono et al. 2001; Radomski et al. 2003; Schweitzer et al. 2008).
This makes the response of the narrow-line flux to changes in the SED complicated.
We plot the slope of BE against the ionization energies of the narrow lines
in the left panel of Fig.~3. We also plot
the Keremedjiev et al. (2009) data for [S{\footnotesize IV}] 10.51$\mu$m:$-0.29\pm0.05$,
[\neii] 12.81$\mu$m:$-0.25\pm0.06$, [\neiii] 15.56$\mu$m:$-0.22\pm0.06$
and [\nev] 14.32$\mu$m:$-0.19\pm0.06$ in purple crosses and the broad-line BE from
Dietrich et al. (2002) in blue rectangles for comparison.
We can see that the narrow-line BE slopes do not correlate with the ionization-energy
( $P_{null} = 0.78$ meaning the two-sided probability that a correlation is not present
is 78\%.) but cluster around -0.2 with a dispersion of $\pm 0.1$.
\niil, \nhal\ and \oiil\ have slopes of $-0.10\pm0.014$, $-0.29\pm0.033$, and $-0.37\pm0.011$ so
they deviate from -0.2 significantly. Possible reasons for this are discussed in Section 4.1.

\subsection{Critical Density -- Slope relationship: No correlation}
It is well known that the NLR is stratified that the high-ionization lines
rise from the inner part of the NLR while
low-ionization lines rise further away (see, for example, Veilleux et al. 1991;
Robinson et al. 1994; Bennert et al. 2006a,b; Kraemer et al. 2009).
The central electron temperature, density, and ionization parameter
are, in general, higher in Seyfert 1s than in Seyfert 2s
(Gaskell 1984; Schmitt 1998; Bennert et al. 2006b).
The lines of lower critical density are not as strong
near the nuclei as the higher critical density lines.
A possible explanation of this is that the change in critical density marks a
transition between difference types of clouds.
To see if this is a factor in the NLR BE we therefore plot the EWs against critical
density in the right panel of Fig.~2 using
the same symbols of panel (a). We again fail to find any correlation
in this plot ($P_{null}=0.82$ using a Spearman rank correlation
analysis).

\section{Discussions}
Using our sample of 3677 radio-quiet AGNs from
SDSS DR4 that have little contamination
by host galaxy, we find that, in contrast with
the BE of broad emission lines, the narrow lines show a nearly
constant BE slope irrespective of
ionization potential or critical density. The \oiil, \niil\ and \nha\ lines show
significantly different BE slopes from other lines.

\subsection{Deviation from constant BE slope}
Although the slope of different lines cluster around -0.2, we do find
some lines that show large deviations.

Firstly,  \niil\ show a much flatter slope ($-0.10\pm0.014$) than other lines ($P < $ 5\% in
two-sided KS test, meaning if the two samples are drawn from the same distribution, we expect to
see a difference as large or larger than we see here only 5 in 100 times.).
This flattening was also found for \nvl, which has an ionization potential
of 77.7 eV but no BE. One plausible explanation of a weaker or absent BE for the nitrogen lines is that
the increase of metallicity in
brighter AGN will enhance nitrogen abundance relative to carbon, oxygen etc., because nitrogen is a
secondary element whose abundance scale as $Z^2$.
This trend could compensate for the decrease of EW in higher-luminosity sources
thus making the BE slope flatter. This explanation is supported by a number of recent studies.
The metallicities of the BLR (Hamann \& Ferland 1993; Nagao et al. 2006a;
Juarez et al. 2009) and NLR (e.g., Nagao et al. 2006b; Matsuoka et al. 2009) are both found to correlate with the
luminosity of the AGN.
So in principal, the $Z_{NLR}$--L relationship could produce
the flattening of the BE of \niil\ we observe here.
However, this explanation has a major problem because \niv] and
\niii], which would be expected to deviate from the
ionization energy--slope relationship for the same reason, lie on
it (Dietrich et al. 2002).
Despite this serious problem to be resolved, Occam's razor suggests that
an enhanced abundance in high-luminosity
AGNs is the simplest and most plausible explanation of the flatter BE slope of
\niil.

The second significant deviation from a BE slope of -0.2 is the steep slopes of \nhal\ and \oiil.
Both of these are star formation (SF)
indicators in star-forming galaxies (Kenicutt et al. 1998, Ho 2005)
and their EW may reach several hundred \AA\ in starburst galaxies.
So even though the AGNs we selected show
no absorption feature in the continuum, the emission lines are
still possible subject to SF contamination because of their large EW.
In low-luminosity sources whose continuum is low, the emission line
from SF may contribute significantly to the total line flux, and thus enhance
the EW.  In type-2 AGNs, the SF contribution of \nhal\ is estimated to be more than
60\% (Brinchmann et al. 2004). The \ha\ emitting region of type-1 AGNs, however, may
have a higher fraction of the emission originating with the AGN (Zhang et al. 2008) but still be
heavily influenced by SF. Furthermore, a variety of studies
have established a correlation
between the strength of AGN activity and star formation in the local
universe ( e.g., Rowan-Robinson 1995; Croom et al. 2002; Netzer et al. 2007, 2009;
Shao et al. 2010). The steepest correlation is $L_{SF}\propto L_{AGN}^{0.8}$
(Netzer 2009) meaning $L_{SF}/L_{AGN}\propto L_{AGN}^{-0.2}$, a BE slope steeper than -0.2.
So it makes sense that the SF in host galaxies would steepen the BE slopes of
\nha\ and \oiil.
However, the \oiil\ and \nhal\ BEs cannot be attributed entirely
to SF--AGN relationship. For a plausible range
of ionization parameters, densities, and ionizing spectra, the intensity
of \oiil\ is proportional to that of \oiiil\ (10\% - 30\%,
Ferland \& Osterbrock 1986; Ho et al. 1993a, 1993b). Since \oiiil\
shows a significant BE, \oiil\ is unlikely to show a radically different
trend. So it is safe to conclude that the \oiil\ and \nhal\ BE could be partly
(but not totally) produced by SF contamination.

A third effect we can see in Fig.~1 is the optical continuum getting bluer
towards higher luminosity. This would leverage the continuum and
lower the EW if the line flux remains unchanged.
This could arise if the NLR sees a filtered spectral-energy distribution (SED)
(Kraemer et al. 1998; Groves et al. 2004a,b) so that a change in the
ionizing spectrum of an AGN would not change the SED the NLR sees much.
The \nevl, \oiil\ and \neiiil\ lines are
most likely to be influenced by a continuum shape difference effect
because of their shorter wavelength.
A possible explanation of the bluer color of the UV-optical SED is the anisotropy of the continuum emission.
Because the accretion disc is optically thick, we will see a dimmer continuum
when viewing it edge-on. (This is the combined result of the $\cos i$ projection effect and the
disk equivalent of ``limb darkening''.)  We could also preferentially be seeing the inner, high-temperature
part of the disc when viewing face-on. These effects will combine to give a higher observed luminosity with a face-on
viewing angle. The reddening of spectrum depends on viewing angle too (Keel 1980;
de Zotti \& Gaskell 1985;  Zhang et al. 2008).
Gaskell et al. (2004) used radio orientations to get AGN reddening curves and found
the continuum shape is profoundly affected by reddening for all but
the bluest AGNs. Because of these effects sources with small inclination would both show large luminosity
and have low reddening.
An alternative explanation of continuum shape difference is host galaxy contamination. Shen et al. (2011) made
composite spectra of different \lfive\ luminosity bins and found that the UV part of all the composite
spectra are similar while the optical part flatten with decreasing luminosity. They interpreted this
trend as due to host galaxy contamination in the optical region of the spectrum in low-luminosity AGNs. Similar
argument is given by Stern \& Laor et al. (2012a).
Even though we have rejected objects with significant stellar light contributions, weak absorption
lines can be spotted on the final composite spectrum of the lowest luminosity bin
because of its extremely high signal to noise ratio.
Our data cannot distinguish different mechanisms that
give rise to the continuum shape difference, and this is beyond the scope
of this paper.
After correcting the deviations listed above,
our conclusion that the slopes of BE for different narrow lines are nearly constant is
further strengthened.

\subsection{Possible causes of the NLR Baldwin effect}
\subsubsection{Softening of the ionizing continuum}
It has been argued on both observational and theoretical grounds that there is a softening of
the ionizing continuum with increasing luminosity
(see, for example, Binette et al. 1989; Netzer, Laor, \& Gondhalekar 1992;
Zheng \& Malkan 1993; Wang \& Lu 1998; Korista et al. 1998).
If this is indeed the case, an important prediction of this
is the ionization potential -- BE slope relationship (Korista et al.
1998). It is successful in explaining the broad line BE, but a major failing is that
it fails to explain the constant BE slope of narrow lines.
It is already known that the NLRs of different AGNs are similar
in the sense that the line ratios show less than 0.5~dex difference
from object to object. (Koski 1978; Veilleux \& Osterbrock
1987; Veilleux 1991a, 1991b, 1991c; Veron-Cetty \& Veron 2000; Dopita et al. 2002;
Gorjian et al. 2007).  Kraemer et al. (2000), Dopita et al. (2002) and Groves et al. (2004a,b)
proposed a NLR model where dust regulates the incident ionizing
spectrum so as to keep the ionization parameter in the NLR constant.
In this model, variation of the SED is filtered by dust in the NLR and thus the continuum shape seen by
the NLR is dominated by the effect of the dust rather than by intrinsic changes in the SED.
This leads to more stable conditions in the NLR. This could
be the cause of the lack of a dependence of the slope of the BE on the ionization energy.

\subsubsection{Luminosity-dependence Covering Factor}
A luminosity-dependent covering factor is a natural explanation of both a BLR and NLR BE.
However, to explain the ionization dependence, the variation in covering factor must have
a different dependence for lines of different ionizations.
This would not be a surprise for the BLR since there is strong radial ionization stratification
and the highest ionization lines are an order of magnitude closer to the black hole than the
lowest ionization lines (see Gaskell 2009 for a review).
Shields, Ferland, \& Peterson (1995) suggested that the broad-line BE could be caused by a luminosity-dependent covering
factor for clouds that are optically-thin to photons with wavelength less than 912\AA.
The optically-thin clouds which have small
column densities, preferentially emit high-ionization lines.
However, comparison of line profiles shows that the BLR BE is due to changes in the {\em low velocity} BLR gas
(Francis et al. 1992) rather than changes in the very broad component.  Furthermore, Snedden \& Gaskell (2007)
argued that optically-thin gas does {\em not} make a
substantial contribution to the BLR.  Nevertheless, given the strong radial ionization stratification of the BLR,
and the decreasing covering factor with ionization (e.g., Francis et al. 1992), an additional luminosity-dependence
of the covering factor could explain the BE and why the high-ionization lines should
show a steeper slope than low-ionization lines.
While this could be consistent with the BLR, it is not obvious how such an explanation can be
reconciled with the lack of an ionization-dependence we find here for the BE.

\subsubsection{A Disappearing NLR?}
The NLR size is correlated with the luminosity of an AGN as $R_{NLR}\propto L^{0.5}$
(Bennert et al. 2002; Schmitt et al. 2003b; Bennert et al. 2006a,b; Greene et al. 2011).
So in high-luminosity AGNs, the NLR may possibly turn from
ionization-bounded to matter-bounded, and the luminosity of narrow lines would
cease increasing with AGN luminosity. This model could in principle explain
part of the BE of NLR (Croom et al. 2002) but it would be too
rash a conclusion that it is the origin of the BE because the
bulk of the NLR emission originates from within the central
few tens or hundreds of pc (Schmitt et al. 2003a) so is unlikely to exceed the
scale of the galactic bulge. Besides, the NLR has no natural size
because it has no definite edge, so some cut in surface brightness or line-ratio
is needed to define the size (Schmitt et al. 2003b; Bennert et al. 2006a,b; Greene et al. 2011).
This makes the interpretation of size complicated.
Netzer et al. (2004) argue that that $R_{NLR}\propto L^{0.5}$ is theoretically sound
yet this relationship must break down for $R_{NLR}$ exceeding a few kpc. They found that high-$z$
AGNs have NLR sizes no larger than 10kpc.
It is also well known that the the NLR is stratified so that the high-ionization lines rise
from the inner part of NLR while low-ionization lines from further away (Veilleux et al. 1995;
Bennert et al. 2006a,b). So the ``disappearing NLR effect'', if exists,
could preferentially influence low-ionization and low-critical density lines.
This again stands against the constant slope we find.


\subsubsection{Continuum Variation}
Because different lines all seem to share a similar BE slope, it is natural to think
that it might be the continuum, rather than the emission lines, which is the cause.
Jiang et al. (2006) proposed that the continuum variation can produce
a weak BE ($\beta=-0.05\pm0.05$) if the light-crossing time
for the region emitting narrow $FeK\alpha$ exceeds the variability timescale for the X-ray
continuum and the amplitude of variability anti-correlates with the luminosity.
Shu et al. (2012) found a strong anti-correlation between the EW of the narrow $FeK\alpha$ line
and $L_X$ ($EW/\langle EW \rangle \propto(L/\langle L \rangle)^{-0.82 \pm 0.10}$ (where $\langle \rangle$ means
time-averaged values) consistent with the X-ray BE
expected in an individual AGN if the narrow-line flux remains
constant while the continuum varies.
For a NLR whose light-crossing time is about $10^3$ yr, the narrow-line luminosity
can safely be assumed to be constant.
The amplitude of variability of the ionizing continuum can be obtained from
the monitoring of AGNs after correction for the constant host galaxy light contribution.
As is well known, the amplitude of variability of an AGN increases with time.
This can be most readily seen from the structure functions of AGNs (the variance as a function of time interval between observations).
We are interested in variability on the longest timescales (timescales similar to the NLR light-crossing time).
We do not, of course, have monitoring on such long timescales, but we can estimate the amplitude from
structure functions.  Cid Fernandez et al. (2000) have presented structure functions for bright AGNs.
These are all consistent with the amplitude of variability rising as the time interval increases to
a characteristic time of a year or a few years and then remaining constant.
Collier \& Peterson (2001) got a similar result for lower-luminosity AGNs (but with a shorter characteristic time)
and also find that the forms of the UV structure functions are similar to the optical ones.
There are indications that the structure function increases gradually on much longer
timescales, but data on this part of the structure function are limited.
Observed variability thus gives a lower limit to variability on the light-crossing timescale of
the NLR.

In order to minimize the host galaxy light contribution one needs to go to as short a
wavelength as possible.  This means using $U$-band observations or space ultraviolet observations.
For example, for NGC~4151 Merkulova (2006) found an {\it observed} peak-to-peak $U$-band amplitude
of a factor of 7.5.  UV variability can be a lot larger.  For example, the peak-to-peak amplitude
at 1300\AA\ for Fairall 9 is a factor of 25 (Koratkar \& Gaskell 1989).
Inspection of the results of long-term UV monitoring of a number of AGNs with the
{\it IUE} satellite (Koratkar \& Gaskell 1989, Koratkar \& Gaskell 1991a,b, Clavel et al. 1991, and O'Brien et al. 1998)
give a median UV peak-to-peak variability of a factor of 7, but the upper quartile
is a factor of 13.  After correction for host-galaxy light, the amplitudes of optical variability
are similarly large.  For example, the NGC 4151 photometry of Lyuty \& Doroshenko (1999)
gives a peak-to-peak amplitude of a factor of about 25 to 30.


Continuum anisotropy (Wang \& Lu 1999) can produce an effect similar to that caused by the
actual variability of the
continuum. It was recently found by Risaliti et al. (2011) that the optically-thick
disc emission responsible for the continuum and isotropic \oiii\
emission will produce the EW(\oiii) distribution very well.
Variability and continuum anisotropy are, in practice,
indistinguishable, so we can consider them together.

We made Monte-Carlo simulations to explore
whether continuum variability can explain the BE slope.
In a flux-limited survey, the luminosity
will show an approximately log-normal distribution.
We generated an artificial sample of sources with a similar distribution of
\lfive\ as the observed AGN sample (0.35~dex
here). We set the EW of the artificial-sources to be the mean
EW of the whole sample, and we add a Gaussian of 0.22~dex to the EW to
account for the intrinsic dispersion.  UV and optical AGN variability is approximately
log-normal -- i.e., it looks normal when plotted in magnitudes (see Gaskell 2004 for a
discussion of log-normal variation of AGNs).
The continuum variation was simulated by adding a Gaussian
with $\sigma=50\%$ (0.18~dex) to the continuum
while the emission line flux was kept unchanged. This is an over-simplification,
but it can help us to gain some insight into
the effect that continuum variability brings.
We generated 4000 sources in each round and measured the slope of
BE using the 4000 sources. An example of the simulation is shown in Fig.~4.
With a peak-to-peak continuum variation of a factor of three, the simulation can produce a BE
slope of $-0.2\pm0.01$ while a factor of six variation produces a BE
slope of $-0.3\pm0.01$. The simulated distribution of EW-\lfive\ can be seen to be very similar
to what is observed.
Obviously, this model will produce a similar slope for
every narrow line if the continuum change with similar amplitude in the
wavelength range we concern. It should be noted that due to variation
in continuum slope, not all lines show the same BE slope, but depends on wavelength.
We assume a constant narrow line flux during the continuum variability. This is an
approximation for the continuum variations on time scales of much shorter than
the light travel time over the NLR because at such short time scale, the NLR has little
response to the continuum variations. For variations on longer time scales, one
needs to properly convolve the continuum variations with the response of emission
lines (transfer function). By considering the latter response, the variations of
equivalent width will be somewhat smaller, but this is equivalent to requiring a
somewhat larger continuum variability amplitude.

Conversely, since continuum variability or the equivalent of continuum variability will inevitably
produce a BE, the BE slope we observe could give an upper-limit on the few $10^2$yrs to $10^3$yrs variability
of AGN optical continuum. If the variability amplitudes exceed the factor of three
a steeper BE will emerge.  This might seem at first sight to be at variance
with the amplitudes of UV variability and structure functions observed, but it must be
remembered that the equivalent width is the ratio of the line flux to the {\it observed}
optical continuum and the observed continuum has a substantial starlight contamination.
An example of how the apparent continuum variability is substantially less than the
real variability is shown in Fig.~4 of Gaskell et al (2008). Host galaxy light limits the
apparent peak-to-peak continuum variability to about a factor of three
(i.e., a rms variability of a few tenths of a magnitude).  This is just what is observed
for the PG AGNs studied by Cid Fernandez et al. Since the structure functions are
relatively flat after a year or so (Cid Fernandez et al. 2000), the variability
effect on the BE should be apparent in a few years. An obvious test of this is to
reobserve AGNs after a few years.


\subsection{Drawbacks}
In summary, the combination of a 50\% variation of \lfive\
and a log-normal distribution of luminosity will naturally produce a -0.20
slope of BE for every narrow line, as we observe. The model we
employ is obviously an oversimplification. To make a more realistic
model, we need a dedicated treatment of sample selection effects and
to make a more realistic assumption of the variability including amplitude,
variation form, and their dependence on wavelength etc, about the light curve.
Despite these simplifications, our results do show that continuum variation will
inevitably produce a similar BE for each narrow line.
Meanwhile, while the model predicts the same BE slope for each
narrow line, the differences in slope are still significant due to
the small error bars. There must therefore be some additional factor at work.
It has been shown in Fig.~3 that the differences
do not correlate with the ionization potential or critical density, so
there must be other factors taking effect.
A deeper exploration of other factors like ionization slope,
NLR geometry and a more realistic model are needed to make progress,
but these are beyond the scope of this paper.

\section{Conclusion}
We have constructed a sample of 3677 $z<0.8$ radio-quiet AGNs
from Sloan Digital Survey Data Release 4 that span three orders of magnitude in
luminosity to explore the relationship between equivalent width of narrow lines
and \lfive. We have computed composite spectra for each \lfive\ bin to
enhance the S/N ratio.
\bha\, and \bhb\ as well as \feii\
were subtracted to get an accurate measurement of narrow emission lines and continuum.
We find that most narrow lines show a
similar BE slope of about -0.2 while the large
deviation of \niil, \oiil, \nhal\ and \nevl\ might
be explained by a metallicity effect,
SF contamination, or the UV-optical continuum difference.
The slope does not show any correlation
with ionization energy and critical density. We propose that
the combination of a 50\% variation of the continuum near 5100\AA\ and a
log-normal distribution of observed luminosity distribution
will naturally produce a -0.2 slope of BE for every narrow line.

\acknowledgements
We thank our referee: Robert Antonucci for inspiring comments that help
to improve the paper significantly. And we thank Brent Groves for helpful discussion on NLR model.
This work is supported by Chinese NSF grants
NSF-10703006, NSF-10973013, NSF-11073019,
and NSF-11233002, the GEMINI-CONICYT Fund of Chile through project N{\degr}32070017 and FONDECYT of Chile
through project N{\degr} 1120957.  Funding for the Sloan Digital Sky Survey (SDSS) has been provided by
the Alfred P. Sloan Foundation, the Participating Institutions,
the National Aeronautics and Space Administration, the National Science
Foundation, the U.S. Department of Energy, the Japanese Monbukagakusho,
and the Max Planck Society. The SDSS is managed by the Astrophysical
Research Consortium (ARC) for the Participating Institutions. The SDSS
web site is http://www.sdss.org/.
\newline

\begin{table*}[h]
\topmargin 0.0cm
\evensidemargin = 0mm
\oddsidemargin = 0mm
\scriptsize 
\caption{Ionization energy, Critical density and
Slope $\beta$ of log EW -- log \lfive\ Table}
\label{corrtab}
\medskip
\vfill
\begin{tabular}{l|c c c c c c c c}
\hline \hline
{Line} &  $\chi_{ion}$  & log $n_c$ &  \lfive\  slope  \\
(1) & (2) & (3) & (4)  \\
\hline \hline
\nevt\    & 97.12   & 7.3  &   -0.31 $\pm$ 0.015  \\
\oiit\    & 13.62   & 3.5  &   -0.37 $\pm$ 0.011  \\
\neiiit\  & 40.96  & 5.5  &   -0.26 $\pm$ 0.020  \\
H$\beta ^n$  & 13.6 & Infinity & -0.28 $\pm$ 0.015  \\
\oiiit\   & 35.11  & 5.8  &   -0.21 $\pm$ 0.016  \\
\oit\     & 0       & 6.3  &   -0.16 $\pm$ 0.030  \\
H$\alpha ^n$  & 13.6 & Infinity &  -0.29 $\pm$ 0.033  \\
\niit\    & 14.5    & 4.82 &   -0.10 $\pm$ 0.014  \\
\siit\    & 10.36    & 2.3  &   -0.20 $\pm$ 0.016  \\
 \hline
\end{tabular}
\medskip
\vfill
{\normalsize ~Columns from left to right: (1)The name of
the line. (2) The ionization energy needed to create the ion. (3) The critical
density of specific line. (4) The log EW -- log \lfive\ slope.}\\
\end{table*}

\begin{figure*}
\begin{center}
\label{fig-1}
\includegraphics[width=16cm]{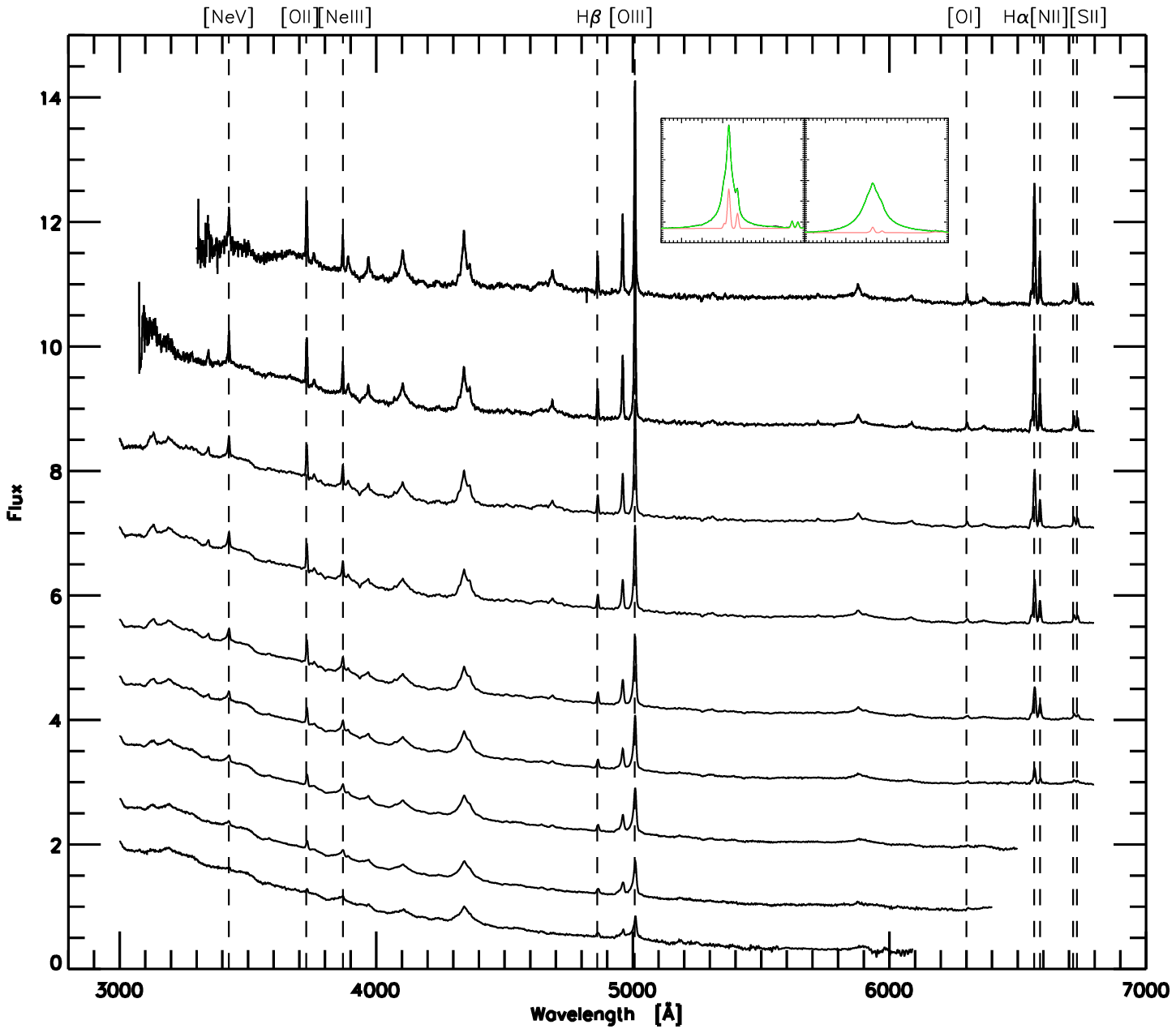}
\caption{Normalized composite spectra with the broad \ha, the broad \hb\ and \feii\
subtracted are shown for luminosity bins of
$\Delta log \lfive= 0.3$ dex, starting from the top with log \lfive=43.35 [$erg s^{-1}$]. 
The spectra are normalized to the [4200\AA,4300\AA] window and shifted
vertically to show the weakening of lines with luminosity more clearly.
The lines we concern with are marked with
dashed lines labeled at the top. The embedded plots show the fitting
results for the \ha+\nii+\sii\ region. The left embedded panel is a
composite spectrum
for log \lfive $\in$ [43.2,43.5] [$erg s^{-1}$]
and the right one is for log \lfive $\in$ [44.4,44.7] [$erg s^{-1}$].}
\end{center}
\end{figure*}

\begin{figure*}
\begin{center}
\label{fig-2}
\includegraphics[width=16cm]{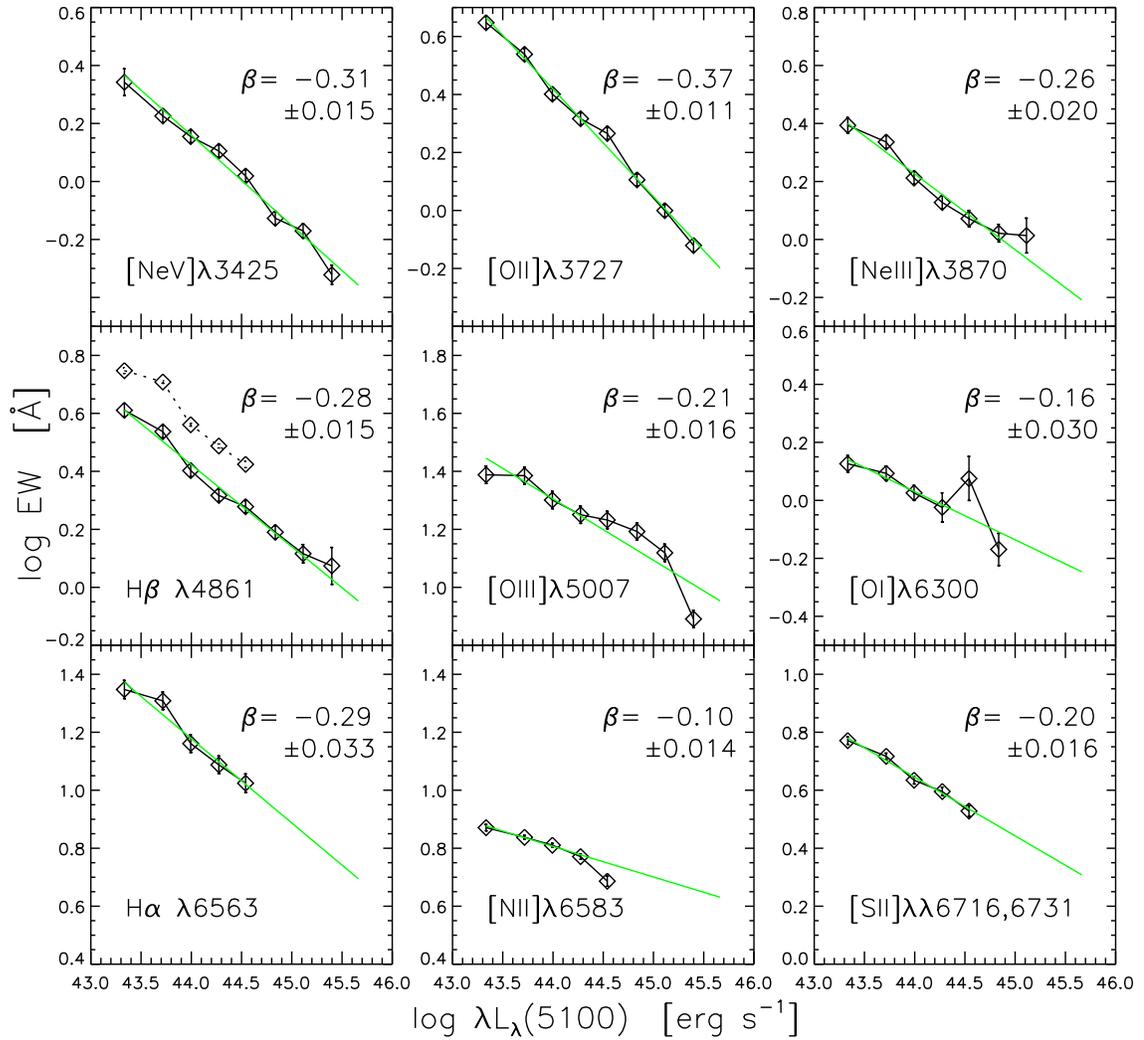}
\caption{Line equivalent widths: $W_{\lambda}$, against 5100\AA\
continuum luminosity: \lfive. We show weighted linear regressions (green lines) for each line
and give the BE slopes as well as their errors in the upper right
corners in each panel. In the narrow \hb\ panel we also include the narrow component
of \ha\ (dashed line) for comparison.}
\end{center}
\end{figure*}

\begin{figure*}
\begin{center}
\label{fig-3}
\includegraphics[width=16cm]{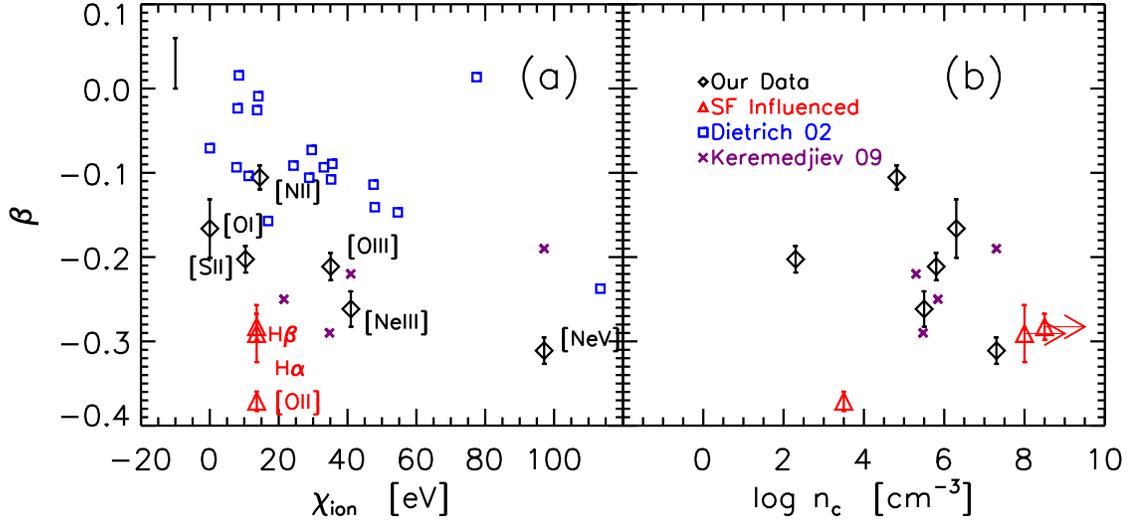}
\caption{Panel (a) BE slopes of different lines against
the ionization energy: $\chi_{ion}$ needed to create the ion. The
diamonds and triangles are our data; the red triangles are lines
that are contaminated by star formation. The data from Dietrich et al. (2002) and
Keremedjiev et al. (2009) are shown with solid blue rectangles and purple
crosses respectively, and the typical error bars are shown in left up corner.
Panel (b)  BE slope against critical density($n_c$) for the same lines
in as in panel (a). The recombination lines are marked with right arrows. }
\end{center}
\end{figure*}

\begin{figure*}
\begin{center}
\label{fig-4}
\includegraphics[width=16cm]{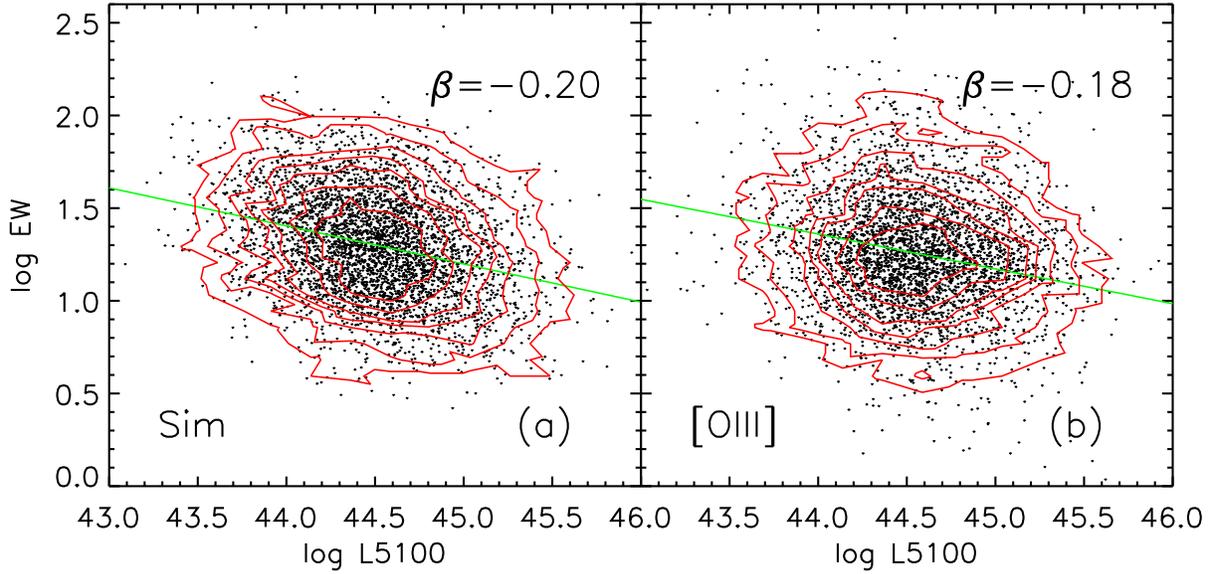}
\caption{The variability-driven BE simulation result.
Panel(a):The simulated BE, the slope of BE is shown on the
right up corner.
Panel(b): The observed EW-log \lfive\ distribution of \oiiil\ from our SDSS
DR4 sample. The slope of BE is shown on the right up corner too.}
\end{center}
\end{figure*}


\clearpage


\begin{thebibliography}{}

\bibitem[Adelman-McCarthy et al.(2006)]{2006ApJS..162...38A}
Adelman-McCarthy, J.~K., et al.\ 2006, \apjs, 162, 38


\bibitem[Antonucci(2002)]{2002apsp.conf..151A} Antonucci, R.\ 2002,
Astrophysical Spectropolarimetry, 151
\bibitem[Antonucci(2012)]{2012arXiv1210.2716A} Antonucci, R.\ 2012,
arXiv:1210.2716

\bibitem[Baldwin(1977)]{1977ApJ...214..679B} Baldwin, J.~A.\ 1977, \apj, 214, 679

\bibitem[Baldwin et al.(1978)]{1978Natur.273..431B}
Baldwin, J.~A, Burke, W.~F., Gaskell, C.~M., \& Wampler, E.~J. 1978, Nature, 273, 431

\bibitem[Baldwin et al.(1989)]{1989ApJ...338..630B} Baldwin, J.~A.,
Wampler, E.~J., \& Gaskell, C.~M.\ 1989, \apj, 338, 630

\bibitem[Bachev et al.(2004)]{2004ApJ...617..171B} Bachev, R., Marziani,
P., Sulentic, J.~W., Zamanov, R., Calvani, M.,
\& Dultzin-Hacyan, D.\ 2004, ApJ, 617, 171

\bibitem[Baskin
\& Laor(2004)]{2004MNRAS.350L..31B} Baskin, A., \& Laor, A.\ 2004, \mnras, 350, L31

\bibitem[Baskin
\& Laor(2005)]{2005MNRAS.358.1043B} Baskin, A., \& Laor, A.\ 2005, \mnras, 358, 1043

\bibitem[Becker et al.(1995)]{1995ApJ...450..559B} Becker, R.~H., White,
R.~L., \& Helfand, D.~J.\ 1995, \apj, 450, 559

\bibitem[Bennert et al.(2002)]{2002ApJ...574L.105B} Bennert, N., Falcke,
H., Schulz, H., Wilson, A.~S., \& Wills, B.~J.\ 2002, \apjl, 574, L105

\bibitem[Bennert et
al.(2006)]{2006A&A...459...55B} Bennert, N., Jungwiert, B., Komossa, S., Haas, M., \& Chini, R.\ 2006, \aap, 459, 55
\bibitem[Bennert et
al.(2006)]{2006A&A...456..953B} Bennert, N., Jungwiert, B., Komossa, S., Haas, M., \& Chini, R.\ 2006, \aap, 456, 953

\bibitem[Binette et al.(1989)]{1989ApJ...343..135B} Binette, L., Prieto,
A., Szuszkiewicz, E., \& Zheng, W.\ 1989, \apj, 343, 135

\bibitem[Clavel et al.(1991)]{1991ApJ...366...64C} Clavel, J., Reichert,
G.~A., Alloin, D., et al.\ 1991, \apj, 366, 64

\bibitem[Cid Fernandes et al.(2000)]{2000ApJ...544..123C}  Cid Fernandes,
R., Sodr{\'e}, L., Jr., \& Vieira da Silva, L., Jr.\ 2000, \apj, 544, 123

\bibitem[Collier \& Peterson(2001)]{2001ApJ...555..775C} Collier, S., \& Peterson, B.~M.\ 2001, \apj, 555, 775

\bibitem[Collin et
al.(2006)]{2006A&A...456...75C} Collin, S., Kawaguchi, T., Peterson, B.~M., \& Vestergaard, M.\ 2006, \aap, 456, 75

\bibitem[Croom et al.(2002)]{2002MNRAS.337..275C} Croom, S.~M., Rhook, K.,
Corbett, E.~A., et al.\ 2002, \mnras, 337, 275

\bibitem[de Zotti
\& Gaskell(1985)]{1985A&A...147....1D} de Zotti, G., \& Gaskell, C.~M.\ 1985, \aap, 147, 1

\bibitem[Denney et al.(2009)]{2009ApJ...692..246D} Denney, K.~D., Peterson,
B.~M., Dietrich, M., Vestergaard, M., \& Bentz, M.~C.\ 2009, \apj, 692, 246

\bibitem[Dibai(1977)]{1977SvAL....3....1D} Dibai, E.~A.\ 1977, Soviet
Astron. Lett., 3, 1

\bibitem[Dietrich et
al.(1999)]{1999A&A...352L...1D} Dietrich, M., Appenzeller, I., Wagner, S.~J., et al.\ 1999, \aap, 352, L1
\bibitem[Dietrich
\& Wilhelm-Erkens(2000)]{2000A&A...354...17D} Dietrich, M., \& Wilhelm-Erkens, U.\ 2000, \aap, 354, 17
\bibitem[Dietrich et al.(2002)]{2002ApJ...581..912D} Dietrich, M., Hamann,
F., Shields, J.~C., Constantin, A., Vestergaard, M., Chaffee, F.,
Foltz, C.~B., \& Junkkarinen, V.~T.\ 2002, ApJ, 581, 912

\bibitem[Dong et al.(2009a)]{2009ApJ...703L...1D} Dong, X.-B., Wang, T.-G.,
Wang, J.-G., et al.\ 2009, \apjl, 703, L1
\bibitem[Dong et al.(2009b)]{2009ASPC..408...83D} Dong, X., Wang, J., Wang,
T., et al.\ 2009, The Starburst-AGN Connection, 408, 83
\bibitem[Dong et al.(2011)]{2011ApJ...736...86D} Dong, X.-B., Wang, J.-G.,
Ho, L.~C., et al.\ 2011, \apj, 736, 86

\bibitem[Dopita et al.(2002)]{2002ApJ...572..753D} Dopita, M.~A., Groves,
B.~A., Sutherland, R.~S., Binette, L., \& Cecil, G.\ 2002, \apj, 572, 753

\bibitem[Ferland
\& Osterbrock(1986)]{1986ApJ...300..658F} Ferland, G.~J., \& Osterbrock, D.~E.\ 1986, \apj, 300, 658

\bibitem[Fine et al.(2008)]{2008MNRAS.390.1413F} Fine, S., Croom, S.~M.,
Hopkins, P.~F., et al.\ 2008, \mnras, 390, 1413

\bibitem[Francis et al.(1992)]{} Francis, P.~J., Hewett, P.~C., Foltz, C.~B., \& Chaffee, F.~H. 1992, ApJ., 398, 476

\bibitem[Gaskell (1984)]{1984ApL....24...43G} Gaskell, C.~M. 1984, Ap.Lett, 24, 43

\bibitem[Gaskell(2004)]{2004ApJ...612L..21G} Gaskell, C.~M.\ 2004, \apjl,
612, L21

\bibitem[Gaskell et al.(2004)]{2004ApJ...616..147G} Gaskell, C.~M.,
Goosmann, R.~W., Antonucci, R.~R.~J., \& Whysong, D.~H.\ 2004, \apj, 616, 147

\bibitem[Gaskell (2008)]{2008RMxAC..32....1G} Gaskell, C.~M.\ 2008, Rev.
Mexicana de Astron. y Astrofis. Conf. Ser., 32, 1

\bibitem[Gaskell(2009)]{2009NewAR..53..140G} Gaskell, C.~M.\ 2009, New Astronomy Reviews,
53, 140

\bibitem[Gaskell \& Klimek(2003)]{2003A&AT...22..661G} Gaskell, C.~M., \& Klimek, E.~S.\ 2003, Astronomical and Astrophysical Transactions, 22, 661

\bibitem[Gorjian et al.(2007)]{2007ApJ...655L..73G} Gorjian, V., Cleary,
K., Werner, M.~W., \& Lawrence, C.~R.\ 2007, \apjl, 655, L73

\bibitem[Greene \& Ho(2005)]{2005ApJ...630..122G} Greene, J.~E., \& Ho, L.~C.\ 2005,
ApJ, 630, 122
\bibitem[Greene et al.(2011)]{2011ApJ...732....9G} Greene, J.~E., Zakamska,
N.~L., Ho, L.~C., \& Barth, A.~J.\ 2011, \apj, 732, 9

\bibitem[Groves et al.(2004)]{2004ApJS..153....9G} Groves, B.~A., Dopita,
M.~A., \& Sutherland, R.~S.\ 2004, \apjs, 153, 9
\bibitem[Groves et al.(2004)]{2004ApJS..153...75G} Groves, B.~A., Dopita,
M.~A., \& Sutherland, R.~S.\ 2004, \apjs, 153, 75

\bibitem[Hamann
\& Ferland(1993)]{1993ApJ...418...11H} Hamann, F., \& Ferland, G.\ 1993, \apj, 418, 11
\bibitem[Hamann
\& Ferland(1999)]{1999ARA&A..37..487H} Hamann, F., \& Ferland, G.\ 1999, \araa, 37, 487

\bibitem[Ho et al.(1993)]{1993ApJ...410..567H} Ho, L.~C., Shields, J.~C.,
\& Filippenko, A.~V.\ 1993, \apj, 410, 567
\bibitem[Ho et al.(1993)]{1993ApJ...417...63H} Ho, L.~C., Filippenko,
A.~V., \& Sargent, W.~L.~W.\ 1993, \apj, 417, 63
\bibitem[Ho(2005)]{2005ApJ...629..680H} Ho, L.~C.\ 2005, \apj, 629, 680

\bibitem[H{\"o}nig et 
al.(2008)]{2008A&A...485L..21H} H{\"o}nig, S.~F., Smette, A., Beckert, T., et
  al.\ 2008, \aap, 485, L21 

\bibitem[Jiang et al.(2006)]{2006ApJ...644..725J} Jiang, P., Wang, J.~X.,
\& Wang, T.~G.\ 2006, \apj, 644, 725
\bibitem[Juarez et
al.(2009)]{2009A&A...494L..25J} Juarez, Y., Maiolino, R., Mujica, R., et al.\ 2009, \aap, 494, L25

\bibitem[Keel(1980)]{1980AJ.....85..198K} Keel, W.~C.\ 1980, \aj, 85, 198

\bibitem[Kennicutt(1998)]{1998ApJ...498..541K} Kennicutt, R.~C., Jr.\ 1998,
\apj, 498, 541

\bibitem[Keremedjiev et al.(2009)]{2009ApJ...690.1105K} Keremedjiev, M.,
Hao, L., \& Charmandaris, V.\ 2009, \apj, 690, 1105

\bibitem[Kollmeier et al.(2006)]{2006ApJ...648..128K} Kollmeier, J.~A.,
Onken, C.~A., Kochanek, C.~S., et al.\ 2006, \apj, 648, 128

\bibitem[Koratkar \& Gaskell (1989)]{} Koratkar, A.~P. \& Gaskell, C.~M. 1989, ApJ., 345, 637

\bibitem[Koratkar \& Gaskell(1991)]{1991ApJ...375...85K} Koratkar, A.~P., \& Gaskell, C.~M.\ 1991, \apj, 375, 85

\bibitem[Koratkar \& Gaskell(1991)]{1991ApJS...75..719K} Koratkar, A.~P., \& Gaskell, C.~M.\ 1991, \apjs, 75, 719

\bibitem[Korista et al.(1998)]{1998ApJ...507...24K} Korista, K., Baldwin,
J., \& Ferland, G.\ 1998, \apj, 507, 24

\bibitem[Koski (1978)]{1978ApJ...223...56K} Koski, A.~T. 1978, ApJ., 223, 56

\bibitem[Kova{\v c}evi{\'c} et al.(2010)]{2010ApJS..189...15K} Kova{\v
c}evi{\'c}, J., Popovi{\'c}, L.~{\v C}.,
\& Dimitrijevi{\'c}, M.~S.\ 2010, \apjs, 189, 15


\bibitem[Kraemer et al.(1998)]{1998ApJ...508..232K} Kraemer, S.~B., Ruiz,
J.~R., \& Crenshaw, D.~M.\ 1998, \apj, 508, 232
\bibitem[Kraemer et al.(2000)]{2000ApJ...531..278K} Kraemer, S.~B.,
Crenshaw, D.~M., Hutchings, J.~B., et al.\ 2000, \apj, 531, 278
\bibitem[Kraemer et al.(2009)]{2009ApJ...698..106K} Kraemer, S.~B., Trippe,
M.~L., Crenshaw, D.~M., et al.\ 2009, \apj, 698, 106

\bibitem[Krolik(2001)]{2001ApJ...551...72K} Krolik, J.~H.\ 2001, \apj, 551,
72

\bibitem[Labiano(2008)]{2008A&A...488L..59L} Labiano, A.\ 2008, \aap, 488, L59

\bibitem[Lawrence(2012)]{2012MNRAS.423..451L} Lawrence, A.\ 2012, \mnras,
423, 451

\bibitem[Lyuty
\& Doroshenko(1999)]{1999AstL...25..341L} Lyuty, V.~M., \& Doroshenko, V.~T.\ 1999, Astronomy Letters, 25, 341

\bibitem[Lu et al.(2010)]{2010MNRAS.404.1761L} Lu, Y., Wang, T.-G., Dong,
X.-B., \& Zhou, H.-Y.\ 2010, \mnras, 404, 1761

\bibitem[Lusso et al.(2012)]{2012MNRAS.425..623L} Lusso, E., Comastri, A.,
Simmons, B.~D., et al.\ 2012, \mnras, 425, 623

\bibitem[Marconi et al.(2008)]{2008ApJ...678..693M} Marconi, A., Axon,
D.~J., Maiolino, R., et al.\ 2008, \apj, 678, 693

\bibitem[Markwardt(2009)]{2009ASPC..411..251M} Markwardt, C.~B.\ 2009, in
Astronomical Data Analysis Software and Systems XVIII, ed. D. A. Bohlender, D.
Durand, \& P. Dowler (San Francisco: ASP), 251

\bibitem[Matsuoka et
al.(2009)]{2009A&A...503..721M} Matsuoka, K., Nagao, T., Maiolino, R., Marconi, A., \& Taniguchi, Y.\ 2009, \aap, 503, 721

\bibitem[Merkulova(2006)]{2006ASPC..360...17M} Merkulova, N.~I.\ 2006,
Astronomical Society of the Pacific Conference Series, 360, 17

\bibitem[Nagao et
al.(2006)]{2006A&A...447..863N} Nagao, T., Maiolino, R., \& Marconi, A.\ 2006, \aap, 447, 863
\bibitem[Nagao et
al.(2006)]{2006A&A...447..157N} Nagao, T., Marconi, A., \& Maiolino, R.\ 2006, \aap, 447, 157


\bibitem[Netzer(1985)]{1985MNRAS.216...63N} Netzer, H.\ 1985, \mnras, 216,
63
\bibitem[Netzer(1987)]{1987MNRAS.225...55N} Netzer, H.\ 1987, \mnras, 225,
55
\bibitem[Netzer et al.(1992)]{1992MNRAS.254...15N} Netzer, H., Laor, A.,
\& Gondhalekar, P.~M.\ 1992, \mnras, 254, 15
\bibitem[Netzer
\& Laor(1993)]{1993ApJ...404L..51N} Netzer, H., \& Laor, A.\ 1993, \apjl, 404, L51
\bibitem[Netzer et al.(2004)]{2004ApJ...614..558N} Netzer, H., Shemmer, O.,
Maiolino, R., Oliva, E., Croom, S., Corbett, E., \& di Fabrizio, L.\ 2004, ApJ, 614, 558
\bibitem[Netzer et al.(2007)]{2007ApJ...666..806N} Netzer, H., Lutz, D.,
Schweitzer, M., et al.\ 2007, \apj, 666, 806
\bibitem[Netzer(2009)]{2009MNRAS.399.1907N} Netzer, H.\ 2009, \mnras, 399,
1907

\bibitem[O'Brien et al.(1998)]{1998ApJ...509..163O} O'Brien, P.~T.,
Dietrich, M., Leighly, K., et al.\ 1998, \apj, 509, 163

\bibitem[Radomski et al.(2003)]{2003ApJ...587..117R} Radomski, J.~T.,
Pi{\~n}a, R.~K., Packham, C., et al.\ 2003, \apj, 587, 117

\bibitem[Rafiee
\& Hall(2011)]{2011ApJS..194...42R} Rafiee, A., \& Hall, P.~B.\ 2011, \apjs, 194, 42

\bibitem[Risaliti et al.(2011)]{2011MNRAS.411.2223R} Risaliti, G., Salvati,
M., \& Marconi, A.\ 2011, \mnras, 411, 2223

\bibitem[Robinson et
al.(1994)]{1994A&A...291..351R} Robinson, A., Vila-Vilaro, B., Axon, D.~J., et al.\ 1994, \aap, 291, 351

\bibitem[Rowan-Robinson(1995)]{1995MNRAS.272..737R} Rowan-Robinson, M.\
1995, \mnras, 272, 737

\bibitem[Schawinski et al.(2007)]{2007MNRAS.382.1415S} Schawinski, K.,
Thomas, D., Sarzi, M., et al.\ 2007, \mnras, 382, 1415

\bibitem[Schmitt, H. R. (1998)]{1998ApJ...506..647S} Schmitt, H R.\  1998 , \apj, 506, 647

\bibitem[Schmitt et al.(2003a)]{2003ApJS..148..327S} Schmitt, H.~R., Donley,
J.~L., Antonucci, R.~R.~J., Hutchings, J.~B.,
\& Kinney, A.~L.\ 2003, \apjs, 148, 327
\bibitem[Schmitt et al.(2003b)]{2003ApJ...597..768S} Schmitt, H.~R., Donley,
J.~L., Antonucci, R.~R.~J., et al.\ 2003, \apj, 597, 768

\bibitem[Schweitzer et al.(2008)]{2008ApJ...679..101S} Schweitzer, M.,
Groves, B., Netzer, H., et al.\ 2008, \apj, 679, 101

\bibitem[Shao et
al.(2010)]{2010A&A...518L..26S} Shao, L., Lutz, D., Nordon, R., et al.\ 2010, \aap, 518, L26

\bibitem[Shakura
\& Sunyaev(1973)]{1973A&A....24..337S} Shakura, N.~I., \& Sunyaev, R.~A.\ 1973, \aap, 24, 337

\bibitem[Shen et al.(2008)]{2008ApJ...680..169S} Shen, Y., Greene, J.~E.,
Strauss, M.~A., Richards, G.~T., \& Schneider, D.~P.\ 2008, \apj, 680, 169
\bibitem[Shen et al.(2011)]{2011ApJS..194...45S} Shen, Y., Richards, G.~T.,
Strauss, M.~A., et al.\ 2011, \apjs, 194, 45

\bibitem[Shields et al.(1995)]{1995ApJ...441..507S} Shields, J.~C.,
Ferland, G.~J., \& Peterson, B.~M.\ 1995, \apj, 441, 507
\bibitem[Shields(2007)]{2007ASPC..373..355S} Shields, J.~C.\ 2007, The
Central Engine of Active Galactic Nuclei, 373, 355

\bibitem[Shu et al.(2012)]{2012ApJ...744L..21S} Shu, X.~W., Wang, J.~X.,
Yaqoob, T., Jiang, P., \& Zhou, Y.~Y.\ 2012, \apjl, 744, L21

\bibitem[Snedden \& Gaskell (2007)]{2007ApJ...669..126S} Snedden, S.~A. \& Gaskell, C.~M. 2007, \apj, 669, 126

\bibitem[Steiner(1981)]{1981ApJ...250..469S} Steiner, J.~E.\ 1981, \apj,
250, 469

\bibitem[Steinhardt
\& Elvis(2010)]{2010MNRAS.402.2637S} Steinhardt, C.~L., \& Elvis, M.\ 2010, \mnras, 402, 2637
\bibitem[Steinhardt
\& Elvis(2010)]{2010MNRAS.406L...1S} Steinhardt, C.~L., \& Elvis, M.\ 2010, \mnras, 406, L1
\bibitem[Steinhardt(2011)]{2011ApJ...738..110S} Steinhardt, C.~L.\ 2011,
\apj, 738, 110

\bibitem[Stern
\& Laor(2012)]{2012MNRAS.423..600S} Stern, J., \& Laor, A.\ 2012, \mnras, 423, 600

\bibitem[Stern
\& Laor(2012)]{2012arXiv1207.5543S} Stern, J., \& Laor, A.\ 2012, arXiv:1207.5543

\bibitem[Stern
\& Laor(2012)]{2012arXiv1210.6394S} Stern, J., \& Laor, A.\ 2012, arXiv:1210.6394

\bibitem[Tomono et al.(2001)]{2001ApJ...557..637T} Tomono, D., Doi, Y.,
Usuda, T., \& Nishimura, T.\ 2001, \apj, 557, 637

\bibitem[Tremonti et al.(2004)]{2004ApJ...613..898T} Tremonti, C.~A.,
Heckman, T.~M., Kauffmann, G., et al.\ 2004, \apj, 613, 898

\bibitem[Vanden Berk et al.(2001)]{2001AJ....122..549V} Vanden Berk, D.~E.,
Richards, G.~T., Bauer, A., et al.\ 2001, \aj, 122, 549

\bibitem[V{\'e}ron-Cetty
\& V{\'e}ron(2000)]{2000A&ARv..10...81V} V{\'e}ron-Cetty, M.~P., \& V{\'e}ron, P.\ 2000, \aapr, 10, 81
\bibitem[V{\'e}ron-Cetty et al.(2004)]{2004A&A...417..515V}
V{\'e}ron-Cetty, M.-P., Joly, M., \& V{\'e}ron, P.\ 2004, \aap, 417, 515

\bibitem[Veilleux
\& Osterbrock(1987)]{1987ApJS...63..295V} Veilleux, S., \& Osterbrock, D.~E.\ 1987, ApJS, 63, 295
\bibitem[Veilleux(1991a)]{1991ApJS...75..357V} Veilleux, S.\ 1991, ApJS,
75, 357
\bibitem[Veilleux(1991b)]{1991ApJS...75..383V} Veilleux, S.\ 1991, ApJS,
75, 383
\bibitem[Veilleux(1991c)]{1991ApJ...369..331V} Veilleux, S.\ 1991, ApJ,
369, 331

\bibitem[Wandel(1999)]{1999ApJ...527..649W} Wandel, A.\ 1999, \apj, 527,
649
\bibitem[Wandel(1999)]{1999ApJ...527..657W} Wandel, A.\ 1999, \apj, 527,
657
\bibitem[Wandel et al.(1999)]{1999ApJ...526..579W} Wandel, A., Peterson,
B.~M., \& Malkan, M.~A.\ 1999, \apj, 526, 579

\bibitem[Warner et al.(2004)]{2004ApJ...608..136W} Warner, C., Hamann, F.,
\& Dietrich, M.\ 2004, \apj, 608, 136

\bibitem[Wang et al.(2009)]{2009ApJ...707.1334W}
Wang, J.-G., et al. 2009, \apj, 707, 1334  

\bibitem[Wang\& Lu(2001)]{2001A&A...377...52W} Wang, T., \& Lu, Y.\ 2001, \aap,
377, 52

\bibitem[Wills et al.(1993)]{1993ApJ...410..534W} Wills, B.~J., Netzer, H.,
Brotherton, M.~S., et al.\ 1993, \apj, 410, 534

\bibitem[Zamorani et al.(1992)]{1992MNRAS.256..238Z} Zamorani, G., Marano,
B., Mignoli, M., Zitelli, V., \& Boyle, B.~J.\ 1992, \mnras, 256, 238

\bibitem[Zhang et al.(2008)]{2008ApJ...685L.109Z} Zhang, K., Wang, T.,
Dong, X., \& Lu, H.\ 2008, \apjl, 685, L109

\bibitem[Zhang et al.(2011)]{2011ApJ...737...71Z} Zhang, K., Dong, X.-B.,
Wang, T.-G., \& Gaskell, C.~M.\ 2011, \apj, 737, 71

\bibitem[Zheng
\& Malkan(1993)]{1993ApJ...415..517Z} Zheng, W., \& Malkan, M.~A.\ 1993, \apj, 415, 517

\bibitem[Zhou et al.(2006)]{2006ApJS..166..128Z} Zhou, H., Wang, T., Yuan,
W., et al.\ 2006, \apjs, 166, 128


\end{thebibliography}
\end{document}